
%
%
%
%
\input harvmac
%
%
%
%
\ifx\answ\bigans
\else
\output={
  \almostshipout{\leftline{\vbox{\pagebody\makefootline}}}\advancepageno
}
\fi
%
%
%
\def\mayer{\vbox{\sl\centerline{Department of Physics 0319}%
\centerline{University of California, San Diego}
\centerline{9500 Gilman Drive}
\centerline{La Jolla, CA 92093-0319}}}
%
%

\def\pyiam{PHY-8958081}

%
%
\def\UCSD#1#2{\noindent#1\hfill #2%
\bigskip\supereject\global\hsize=\hsbody%
\footline={\hss\tenrm\folio\hss}}
%
%
\def\abstract#1{\centerline{\bf Abstract}\nobreak\medskip\nobreak\par #1}
%
%
%
%
\edef\tfontsize{ scaled\magstep3}
 \tfontsize  \tfontsize
 \tfontsize \font\titlei=cmmi10 \tfontsize
\font\titleis=cmmi7 \tfontsize \font\titleiss=cmmi5 \tfontsize
\font\titlesy=cmsy10 \tfontsize \font\titlesys=cmsy7 \tfontsize
\font\titlesyss=cmsy5 \tfontsize  \tfontsize
\skewchar\titlei='177 \skewchar\titleis='177 \skewchar\titleiss='177
\skewchar\titlesy='60 \skewchar\titlesys='60 \skewchar\titlesyss='60
%
%
%
%
%
\def\inv{^{\raise.15ex\hbox{${\scriptscriptstyle -}$}\kern-.05em 1}}
\def\lbar{{\lower.35ex\hbox{$\mathchar'26$}\mkern-10mu\lambda}} 

%
%
%
%
\def\slash#1{\rlap{$#1$}/} 
\def\dsl{\,\raise.15ex\hbox{/}\mkern-13.5mu D} 
\def\delsl{\raise.15ex\hbox{/}\kern-.57em\partial}
\def\Ksl{\hbox{/\kern-.6000em\rm K}}
\def\Asl{\hbox{/\kern-.6500em \rm A}}
\def\Dsl{\hbox{/\kern-.6000em\rm D}} 
\def\Qsl{\hbox{/\kern-.6000em\rm Q}}
\def\gradsl{\hbox{/\kern-.6500em$\nabla$}}
%
%
\def\lspace{\ifx\answ\bigans{}\else\qquad\fi}
\def\lbspace{\ifx\answ\bigans{}\else\hskip-.2in\fi} 
%
%
\def\boxeqn#1{\vcenter{\vbox{\hrule\hbox{\vrule\kern3pt\vbox{\kern3pt
        \hbox{${\displaystyle #1}$}\kern3pt}\kern3pt\vrule}\hrule}}}
%
%
\def\mbox#1#2{\vcenter{\hrule \hbox{\vrule height#2in
\kern#1in \vrule} \hrule}}
%
%
%
%

\def\CM{{\cal M}}  \def\CO{{\cal O}}

%
%
%
%
%
\def\del{\partial}

\def\bar#1{\overline{#1}}

\def\bra#1{\left\langle #1\right|}
\def\ket#1{\left| #1\right\rangle}
\def\abs#1{\left| #1\right|}

\def\darr#1{\raise1.5ex\hbox{$\leftrightarrow$}\mkern-16.5mu #1}

%
%
\def\frac#1#2{{\textstyle{#1\over #2}}} 
%
%
%
%

\def\Tr{\mathop{\rm Tr}}

%
%
%
%

%
%
\def\ltap{\ \raise.3ex\hbox{$<$\kern-.75em\lower1ex\hbox{$\sim$}}\ }
\def\gtap{\ \raise.3ex\hbox{$>$\kern-.75em\lower1ex\hbox{$\sim$}}\ }
\def\gl{\ \raise.5ex\hbox{$>$}\kern-.8em\lower.5ex\hbox{$<$}\ }
\def\roughly#1{\raise.3ex\hbox{$#1$\kern-.75em\lower1ex\hbox{$\sim$}}}
%
%
\def\ie{\hbox{\it i.e.}}

\def\np#1#2#3{{Nucl. Phys. } B{#1} (#2) #3}
\def\pl#1#2#3{{Phys. Lett. } {#1}B (#2) #3}

\def\physrev#1#2#3{{Phys. Rev. } {#1} (#2) #3}

\relax

\def\a{a_0}
\def\aa{\a\, }

\def\as{{\alpha_s}}
\def\lqcd{\Lambda_{{\rm QCD}}}

\def\D{{\rm D}}

\def\CO{{\cal O}}
\def\mev{\rm MeV}

\def\lta{\ \hbox{\raise.55ex\hbox{$<$}} \!\!\!\!\!
\hbox{\raise-.5ex\hbox{$\sim$}}\ }
\def\gta{\ \hbox{\raise.55ex\hbox{$>$}} \!\!\!\!\!
\hbox{\raise-.5ex\hbox{$\sim$}}\ }

\def\mayer{\vbox{\sl\centerline{Department of Physics}
\centerline{9500 Gilman Drive 0319}
\centerline{University of California, San Diego}
\centerline{La Jolla, CA 92093-0319}}}
\noblackbox
\vskip 1.in
\centerline{{\titlefont{Semileptonic $B_c$ Decay}}}
\medskip
\centerline{{\titlefont{and Heavy Quark Spin Symmetry}}}
\bigskip
\vskip .5in
\centerline{Elizabeth Jenkins, Michael Luke, Aneesh V. Manohar
and Martin J. Savage}
\bigskip\medskip
\mayer
\bigskip\bigskip
\abstract{
Semileptonic decay of the $B_c$ meson is studied in the heavy quark
limit. The six possible form factors for $B_c
\rightarrow B_s (B^0),B_s^* (B^{*0})$
semileptonic decay
are determined by two invariant functions.  Only one of these
functions
contributes at zero recoil, where it is calculable
to lowest order in an operator product expansion
in terms of the meson decay
constant $f_B$ and the $B_c$ wavefunction.
A similar result is found for $B_c \rightarrow D^0,D^{*0}$ and
for $B_c\rightarrow\eta_c,J/\psi$ semileptonic decay for a restricted kinematic
region. Semileptonic $B_c$ decay provides a means for determining
the KM mixing angle $|V_{ub}|$.
}
\vfill
\UCSD{UCSD/PTH 92-13}{April 1992}

\newsec{Introduction}

The $B_c$ meson provides a unique probe of both strong and weak
interactions.  Unlike quarkonium systems which can decay strongly and
electromagnetically, the $B_c$ can only decay weakly and thus is
relatively long-lived.  The $c$ and $b$ quark lifetimes are similar
because of the small mixing angle $V_{cb}$, so $B_c$ decay proceeds
through either quark at comparable rates. In this work, we study
semileptonic weak decay of the $B_c$ meson by exploiting heavy quark
spin symmetry. We will compute the weak decay amplitude for
$B_c\rightarrow B_s$ in terms of the $B_s$ meson decay constant
$f_{B_s}$. The two are related because the $B_c$ can be treated as a
pointlike meson in the limit that the $b$ and $c$ quark masses are much
larger than $\lqcd$.  The weak decay of the $c$ quark produces a state
which has a $\bar b$ and $s$ quark at the same point in space
(to lowest order in an operator product expansion); the
amplitude for this state to turn into a $B_s$ is $f_{B_s}$. A similar
argument allows us to compute the amplitudes for
$B_c\rightarrow D^0, D^{*0}$ and $B_c\rightarrow \eta_c, J/\psi$
semileptonic decay.

The standard application of heavy quark symmetry is to hadrons
containing a single heavy quark.  We must take some care in defining the
heavy quark effective theory when dealing with a system with two heavy
quarks.  It is well known that the static theory gives rise to severe
infrared divergences for diagrams involving two heavy quarks with the
same velocity \ref\irdiv{B.~A.~Thacker and G.~P.~Lepage,
\physrev{D43}{1991}{196}.}.  These divergences are regulated by the
kinetic energy term $\bar h_Q(\D^2/2m_Q)h_Q$; even though
this term is higher order in
$1/m_Q$, it may not be neglected in the $m_Q\rightarrow\infty$ limit.
The kinetic energy term is different for $b$ and $c$ quarks, and breaks the
heavy flavor symmetry. Physically, this is just the statement that the
dynamics of heavy-heavy bound states is determined by balancing the
kinetic and potential energies of the quarks; the $\Upsilon$ is not the
same size as the $J/\psi$, and we cannot use the heavy flavor symmetry
to relate these two states.  The kinetic term in the effective
Lagrangian breaks the heavy flavor symmetry, but it does not break the
heavy quark spin symmetry. Thus we can still derive relations for
hadrons with two heavy quarks using heavy quark spin symmetry.

\newsec{Spin Symmetry}

The invariance of the effective Lagrangian under individual spin
rotations on the $b$ and $c$ quarks allows us to relate the form factors
for vector and axial vector currents between the $B_c$ and
various pseudoscalar and vector
mesons in the same way as for heavy-light systems
\ref\wis{N. Isgur and M.B. Wise, \pl {232}{1989}{113}
\semi \pl {237}{1990}{527}.}.

Let us first consider the semileptonic decays $B_c\rightarrow B_s(B^0)
e^+\nu$ or $B_c\rightarrow B^*_s(B^{*0}) e^+\nu$ in which the the $B_c$
decays into a $B$ meson containing a light quark.  These decays
correspond to
the semileptonic weak decay of the charm quark into a light $s$ or $d$
quark.  Since the mass of the $\bar b$ quark is much greater than that
of the $c$ quark, the energy released in the decay of the $c$ quark is
much smaller than $m_b$, and the $\bar b$ quark is not deflected.  Thus
the velocity of the final meson is the same as the velocity of the
initial meson. The initial momentum of the $B_c$ is $p^\mu=m_{B_c}
v^\mu$, and the final momentum of the $B_a$ is $p'^{\mu} = m_{B_a} v^\mu
+ q^\mu$, where $q$ is a small residual momentum.  The final $B_a$ is on
shell, so $q\cdot v=\CO(1/m_B)$. The momentum transfer to the lepton system is
\eqn\kqreln{
k^\mu=p^\mu-p'^\mu=(m_{B_c}-m_{B_a})v^\mu - q^\mu.  }

Heavy quark spin symmetry implies that the pseudoscalar $B_c$ meson is
degenerate with the vector $B_c^*$ meson.  The consequences of spin
symmetry for hadronic matrix elements
may be derived using the commutation relations of Isgur and
Wise \wis, or more compactly using the well-known trace
formalism \ref\traform{A.~F.~Falk, H.~Georgi, B.~Grinstein and
M.~B.~Wise,
\np{343}{1990}{1}\semi J.~D.~Bjorken, talk given at Les Rencontres de la
Valle d'Aoste La Thuile, Aosta Valley, Italy, March 1990, SLAC preprint
SLAC-PUB-5278 (1990).}.  The lowest-lying $\bar b c$ bound
states are represented by a $4 \times 4$ matrix
\eqn\bch{H^{(c\bar b )} = {{(1+\slash v)}\over 2}\left[
B_c^{*\mu}\gamma_\mu-B_c\gamma_5\right] {{(1-\slash v) }\over 2}\ \ \ ,}
where $B_c$ and $B_c^*$ annihilate pseudoscalar and vector meson $\bar b
c$ bound states of velocity $v$, respectively.  A subscript $v$ on
the heavy meson fields has been suppressed.  Under spin symmetries on the
heavy quark and antiquark, the heavy meson field transforms as
\eqn\trans{H^{(c\bar b )}\rightarrow S_c\ H^{(c\bar b )}\
S_{ b}^\dagger\ \ .} An analogous definition for $H^{(c \bar c)}$
describes the ($\eta_c$, $J/\psi$) spin multiplet.  The spin multiplet
for the $B_a$ and $B_a^*$ is given by
\eqn\bsh{H^{(\bar b)}_a =
\left[B_a^{*\mu}\gamma_\mu-B_a\gamma_5\right]
{{(1 - \slash v)}\over 2} \ \ ,} where the subscript $a=1,2,3$ (or
$u,d,s$) is an $SU(3)_V$ flavor index.  The field $H^{(\bar b)}_a$ is a
doublet under heavy quark spin symmetry and a $3$ under flavor $SU(3)_V$
symmetry \ref\mwise{M.B. Wise, CALT-68-1765 (1992)
\semi G. Burdman and J. Donoghue, UMHEP-365 (1992).},
\eqn\tranu{
H^{(\bar b)}_a\rightarrow \left(U H^{(\bar b)}\right)_a S^\dagger_b\ .}
Note that the pseudoscalar and vector meson fields $B$ and $B^*$ have
dimension $3/2$ because they contain factors of $\sqrt{m_B}$ and
$\sqrt{m_{B^*}}$ relative to the standard normalization for scalar and
vector fields.

The amplitudes for semileptonic $B_c$ decay to $B_a$ and $B_a^*$ are
determined by the matrix elements of the weak hadronic current $\bar q_a
\gamma_\mu (1 - \gamma_5) c$ between the meson states.  The most general
form for the matrix element of the current
which respects the heavy
quark spin symmetry is
\eqn\invt{\bra{B_a^{(*)},v,q}\bar q_a\Gamma c\ket{B_c,v^{\vphantom{(*)}}} =
-\sqrt{m_{B_c} m_{B_a}}\Tr\left(\bar H^{(\bar b)}_a
\ \Omega(v,\a q)\ \Gamma\ H^{(c\bar b)}\right),}
where
\eqn\oms{\Omega(v,\a q)=\Omega_1(\a q)+ \aa\Omega_2(\a q)\ \slash{q},}
is the most general Dirac matrix that can be written in terms of the
vectors $q$ and $v$ (recall that $q\cdot v$=0).
Terms with factors of $\slash v$ can be omitted because of the
identities
\eqn\videntity{
\slash v H^{(c\bar b)} = H^{(c\bar b)},\quad H^{(c\bar b)} \slash v
= - H^{(c\bar b)}, \quad \slash v H_a = H_a,\quad H_a\slash v = -H_a .}
Note that the factor of $\Gamma$ multiplying $H^{(c\bar b)}$ in Eq.~\invt\
is required by the heavy quark spin symmetry on the $c$ quark.
The radius
of the $B_c$ meson, $\a$, is the
typical scale for the variation of the form factors (as will be shown in
the next section).  For a Coulomb bound state, $\a^{-1}\sim
\as\left(\a^{-1}\right)m_c$;  the linear confining term in the potential
makes the state
somewhat smaller than this estimate.  In our case, $\a(B_c)\simeq
\a(J/\psi)\simeq(500\ \mev)^{-1}$ \ref\rosner{C.~Quigg and J.~L.~Rosner,
\physrev{D23}{1981}{2625}.}.

Explicit evaluation of Eq.~\invt\ gives
\eqn\ppv{\eqalign{
\langle B_a,v,q|V_\mu|B_c,v\rangle = &
\sqrt{2 m_{B_c}\  2m_{B_a}}\,\left[\Omega_1 \,v_\mu + \ \aa\Omega_2
\ q_\mu\right]
 \, ,\cr
\langle B_a^*,v,q|V_\mu|B_c,v\rangle = & -i\sqrt { 2m_{B_c}\ 2m_{B^*_a}}
\ \aa\Omega_2\ \epsilon_{\mu\nu\alpha\beta} \ \epsilon^{*\nu} q^\alpha
v^\beta\, , \cr
\langle B_a^*,v,q|A_\mu|B_c,v\rangle =  &
\sqrt{ 2m_{B_c}  \ 2m_{B^*_a}}\,\left[\Omega_1\, \epsilon_\mu^*
+ \aa\Omega_2\ \epsilon^*\cdot q\ v_\mu\right]\, ,}} where $V_\mu$ and
$A_\mu$ refer to the vector and axial vector currents $\bar q_a
\gamma_\mu c$ and $\bar q_a \gamma_\mu \gamma_5 c$, respectively, and
$\epsilon_\mu$ is the polarization vector of the $B_a^*$.  The form
factor $\Omega_2$ is irrelevant for semileptonic $B_c\rightarrow
B_s(B^0)$ decay because the contribution of $\Omega_2$ to the
decay amplitude will be proportional to the lepton mass.
In addition, $\Omega_2$ does not contribute to decay amplitudes
at zero recoil, $q=0$.
Note that the dimensionless functions $\Omega_i(\a q)$ are
independent of the light quark flavor index in the $SU(3)_V$ limit.
Thus, the ratio
of KM mixing angles $| V_{cs}/ V_{cd} |$ can be extracted from
comparison
of $B_c$ semileptonic decay to $B_s, B_s^*$ and $B^0, B^{*0}$.
Leading $SU(3)_V$-violating
light quark flavor dependence of the form factors
may be estimated in chiral
perturbation theory
\ref\js{E. Jenkins and M.J. Savage, UCSD/PTH 92-07.}.

A similar analysis applies to the decays $B_c
\rightarrow D^0$ and $B_c \rightarrow D^{*0}$ in which the $\bar b$
quark decays to a $\bar u$.  In this case, however, the light antiquark
will typically recoil with momentum comparable to or larger than the $c$
quark mass.  In order for the final meson to be bound, there must be a
correspondingly large momentum transfer to the spectator $c$ quark, and
the effective theory breaks down.  Higher dimensional operators which we
have neglected will be of order $q/m_c$ and will dominate for large
momentum transfer to the light degrees of freedom.  Our results are thus
valid only for $q\ll m_c$, \ie\ near the zero recoil point.  With this
caveat, the analysis proceeds exactly as before. The amplitudes can be
written in terms of two invariant functions $\Sigma_1(\a q)$ and
$\Sigma_2(\a q)$,
\eqn\ppvb{\eqalign{
\langle D^0,v,q|V_\mu|B_c,v\rangle = &
\sqrt{2 m_{B_c} 2 m_{D}}\,\left[\Sigma_1  \,v_\mu + \aa\, \Sigma_2 \,
q_\mu\right] \, ,\cr
\langle D^{*0},v,q|V_\mu|B_c,v\rangle = & -i\sqrt {2 m_{B_c} 2 m_{D^*}}
\,\aa\, \Sigma_2\, \epsilon_{\mu\nu\alpha\beta} \epsilon^{*\nu}
q^\alpha v^\beta\, , \cr
\langle D^{*0},v,q|A_\mu|B_c,v\rangle =  &
\sqrt{2 m_{B_c} 2 m_{D^*}}\,\left[\Sigma_1\, \epsilon_\mu^*
+\aa\, \Sigma_2 \, \epsilon^*\cdot q \ v_\mu\right].}}
Measurement of this decay provides a means of determining the KM
angle $| V_{ub} |$.

Finally, we analyze the semileptonic decays of the ${B}_c$ to the
charmonium mesons $\eta_c$ and $\psi$.  Once again, the momentum
transfer to the produced $\bar c$ quark
may be large and our results are only
valid near the zero recoil point.  In this case, however, there is an
additional spin symmetry of the produced
antiquark, which forbids a form factor proportional to $\slash q$.
Thus, the matrix elements for the semileptonic decay of $B_c$ to
$\eta_c$ and $\psi$ near zero recoil
are determined by a single function $\Delta (\a q)$:
\eqn\ppvv{\eqalign{
\langle \eta_c, v, q |V_\mu| B_c, v \rangle = &
\sqrt{2 m_{B_c}2 m_{\eta_c}}\ \Delta\ v_\mu, \cr
\langle \psi, v, q|A_\mu| B_c, v\rangle = &
\sqrt{2 m_{B_c}2 m_{\psi}}\ \Delta \ \epsilon_\mu^*,}}
where $V_\mu$ and $A_\mu$ refer to the vector and axial vector currents
$\bar b \gamma_\mu c$ and $\bar b \gamma_\mu \gamma_5 c$, respectively,
and $\epsilon_\mu$ is the polarization vector of the $\psi$.

\newsec{The Scale of Variation of Form Factors}

The invariant tensors in Eq.~\invt\ are multiplied by dimensionless form
factors $\Omega_i(\a q)$.
In this section, we explain
in greater detail why the scale of variation of the form factors
is set by $\a$, the radius of the $B_c$
bound state.
Before discussing the case of interest, it is useful to first
consider the scale of variation of form factors for two different
circumstances --- the matrix elements for a heavy quark current between
two heavy-light mesons  and the matrix elements for a light quark
current between two heavy-light mesons.  Let us first analyze the
well-known example of the matrix elements of a heavy quark current
between meson states containing a single heavy quark.  For concreteness,
consider the matrix elements of the current
$\bar b \Gamma b$ between $B$ and $B^*$ mesons.  The matrix elements
can be evaluated using the trace formalism,
\eqn\bdtrace{
\bra{B,v'} \bar b \Gamma b \ket{ B,v}=
 m_{B}\ \Tr
\left( \bar H_{v'}^{(\bar b)}\ \Omega(v,v')\ H_v^{(\bar b)}\ \Gamma
\right)\ ,}
where $v$ and $v'$ are the velocities of the initial and final meson
fields, respectively.
The Dirac matrix coupling the heavy quark spin indices in the trace must
be $\Gamma$ by the spin symmetry. The matrix $\Omega$ coupling the light
quark indices is not constrained, and is the most general possible Dirac
matrix that can be constructed out of $v$ and $v'$.
{}From
Eq.~\videntity, it follows that there is only one possible invariant
for $\Omega$,  the
Isgur-Wise function $\xi(v\cdot v')$.  This dimensionless
nonperturbative function is the form factor for the light
degrees of freedom in the heavy-light mesons.  The light degrees
of freedom of a $B$ meson with definite velocity $v$
carry a momentum which is typically
of order $\lqcd v$.  Thus, the momentum transfer
to the light degrees of freedom in the above transition is of
order $\lqcd(v-v')$. Since hadronic form factors for the light degrees of
freedom vary on the momentum scale $\lqcd$, the
variation of the Isgur-Wise function is controlled by
$\lqcd(v-v')/\lqcd=v-v'$. Thus the Isgur-Wise function $\xi(v\cdot v')$
varies on the scale over which $v\cdot
v'$ changes by order one.\foot{The $B^*\rightarrow B\gamma$
electromagnetic
transition amplitude due to the $b$ quark current $\bar b\gamma^\mu b$
can be computed from Eq.~\bdtrace\ with the substitution
$\Gamma=\gamma^\mu$.  The amplitude
vanishes; thus the $b$ quark magnetic moment transition is a $1/m_b$
effect.}

The scale of variation of form factors is different for the matrix
elements of a light quark current
between the same states.  In the following,
we will assume that the momentum transfer of the transition is
small compared with the mass of the heavy quark, so that
the velocity of the $B$ meson is not changed.
The matrix elements of the light quark current $\bar d \gamma_{\mu} d$
are given by
\eqn\bbtrace{
\bra{ B,v,q} \bar d \Gamma d \ket{ B,v,0}=
{m_{B}}\ \Tr
\left(\bar H_{v}^{(\bar b)}\ \Omega_\Gamma(v,q) \ H_v^{(\bar b)}\right),
}
where the states are described by both a velocity $v$ and a
residual momentum $q$ \ref\georgi{H. Georgi, \pl{240}{1990}{447}.}
\ref\dgg{M.~J.~Dugan, M.~Golden and B.~Grinstein,
Harvard preprint HUTP-91/A045 (1991).}.  The Dirac
matrix coupling the heavy quark indices is the identity matrix
because of the heavy quark spin symmetry. The Dirac matrix
$\Omega_\Gamma$ coupling the light quark indices is the most general
Dirac matrix which transforms as $\Gamma$ under Lorentz transformations.
For $\Gamma=\gamma^\mu$, we find\foot{The $\Omega_1$ form
factor is the electric coupling, and the $\Omega_2$ form factor is the
magnetic coupling.
The $\Omega_2$ form factor gives a $B^*\rightarrow B\gamma$ transition
amplitude that is not suppressed by powers of $1/m_B$, and corresponds
to a light quark magnetic moment transition in a quark model.}
\eqn\bbffactor{
\Omega_{\Gamma}=\Omega_1(q^2/\lqcd^2)\ v^\mu
+ \Omega_2(q^2/\lqcd^2)  \ \sigma^{\mu\nu}
q_\nu/\lqcd, }
where we have used
current conservation in writing Eq. \bbffactor.
In contrast to the first example, the momentum transfer
to the light degrees of
freedom is $q$, so that the scale of variation of the form factors is
now $q/\lqcd$.
The form factors $\Omega_i$ in Eq.~\bbffactor\ have
a variation on the scale $p\cdot p'\sim \lqcd^2 $ (or
$v\cdot v'\sim \lqcd^2/m_B^2$) instead of the scale
$v\cdot v'\sim1$ for
the Isgur-Wise function $\xi(v\cdot v')$.

We now consider the scale of variation for the form factors found
in Sect.~2.
The $B_c\rightarrow B_s$ transition amplitude is an example of a matrix
element of an operator containing both a light quark and a heavy quark.
The matrix elements of the current $\bar s \Gamma c$ are given by
\eqn\hlmatrix{
\bra{B_s,v,q} \bar s \Gamma c \ket{B_c,v,0}=
-\sqrt{ m_{B_c}  m_{B_s} }\ \Tr\left(
\bar H_{s \ v}^{(\bar b)}\ \left[\Omega_1 + \aa\Omega_2\ \slash q\right]\
\Gamma \ H_v^{(c \bar b)}\right) ,
}
where the matrix $\Gamma$ multiplies $H_v^{(c \bar b)}$ on the $c$
quark index because of the $c$ quark spin symmetry, and $\Omega_1 +
\aa\Omega_2\slash q$ is the most general possible scalar matrix,
and multiplies the light quark index.
In the limit $m_b\gg m_c\gg\lqcd$, the
addition of momentum $q$ does not change the velocity of the meson.
The scale of variation of the form factors is controlled by the
size of the $B_c$ bound state.  The
matrix element Eq.~\hlmatrix\ measures the overlap of the $c$ quark
distribution in the $B_c$, the $s$ quark distribution in the $B_s$, and
$e^{iq\cdot x}$. Equivalently, it measures the overlap of the $s$ quark
distribution in the $B_s$ with the $c$ quark distribution in the $B_c$
shifted by momentum $q$. The width of the momentum distribution of the
$s$ quark is $\lqcd$, and the width of the momentum distribution of the
$c$ quark is of order the inverse radius of the
$B_c$ bound state, $\a^{-1}\gg \lqcd$.
Thus a shift in the $c$ quark momentum distribution
by an amount $q \ll \a^{-1}$ does not affect the overlap amplitude.
Consequently,
the scale of variation of the form factors is $\a^{-1}$,
not $\lqcd$.

\newsec{The Zero Recoil Limit}

The weak currents $\bar s\Gamma c$, $\bar d \Gamma c$, $\bar b \Gamma u$
and $\bar b\Gamma c$ do not generate
symmetries of the effective theory, so their matrix elements at zero
recoil cannot be normalized by symmetry considerations.  However, in the
limit $1/\a\gg\lqcd$ in which the
$B_c$ is
pointlike on the hadronic scale
$\lqcd$, it is possible to calculate
these matrix elements at zero recoil in terms
of heavy-heavy bound state wavefunctions and the meson
decay constants $f_{B}$ and $f_D$.

We begin by considering semileptonic $B_c \rightarrow B_s$ decay.
In the $m_b\rightarrow\infty$ limit, the kinematics of $B_c\rightarrow
B_s$ decay is analogous to that of neutron $\beta$-decay in that the entire
energy of the decay is transferred to the lepton system.
In the
rest frame of the $B_c$, $k^0=E_\ell+E_\nu=m(B_c)-m(B_s)$, where $k$ is
defined in Eq.~\kqreln, or equivalently $q^0=0$. (The
recoil energy of the $B_s$ is of order ${\vec q}^{\,2}/m_b$.)  Thus the
hadronic form factors only depend on $\vec q$, and the zero recoil point is
$\vec q=0$.

The calculation of the form factor at zero recoil proceeds
as follows.
The initial $B_c$ state is written as\foot{States with a subscript HQ are
normalized to $v^0$ rather than to $2E$.}
\eqn\wave{\ket{B_c, v}_{{\rm HQ}} = \int d^3x\ \Psi (x) \left[
\overline{c}^{(+)}_v(x){{(1+\slash v)}\over 2}i\gamma_5{{(1-\slash
v)}\over 2} b^{(-)}_v(0)\right]|0\rangle\ \ \ ,}
where $b^{(-)}_v(0)$
creates a $\bar b$ quark with velocity $v$ at the origin,
$\overline{c}^{(+)}_v(x)$ creates a $c$ quark with velocity $v$ at
the
point $x$, and $\Psi(x)$ is the wavefunction of the $B_c$. The superscripts
$(+)$ and $(-)$ refer to the $\slash v$
eigenvalue,
\eqn\veigen{
\slash v c_v^{(+)} = + c_v^{(+)},\qquad
\slash v c_v^{(-)} = - c_v^{(-)},
}
and similarly for $b_v^{(\pm)}$.
To compute the $\Omega_1$ form factor in Eq.~\ppv\ for $B_c\rightarrow B_s$
semileptonic decay, we consider the matrix element of the vector current
between the $B_c$ and $B_s$ at finite three momentum transfer $\vec q$.
\eqn\mint{\eqalign{
\CM^{\mu}(\vec q)&=
\int d^3z\ e^{i\vec q\cdot\vec z}\,
{}_{{\rm HQ}}\langle B_s, v|V^\mu|B_c, v\rangle_{{\rm HQ}} \cr &=
\int d^3z\ e^{i{\vec q\cdot \vec z}}\,
{}_{{\rm HQ}}\langle B_s, v| \overline{s}(z)\gamma^\mu
c_v(z)|B_c, v\rangle_{{\rm HQ}}  \ ,\cr}}
where ${\vec q}$ is the three-momentum transfer to the leptonic system
in
the decay.
Inserting Eq.~\wave\ into Eq.~\mint, and using heavy field
contractions,
\eqn\hqcont{
\bra{0} c_v(x)\bar c_v^{(+)}(y) \ket{0} = {1+\slash v\over 2}\delta(x-y),
}
yields
\eqn\minty{
\CM^\mu(q)=
i\int d^3x\ e^{i{\vec q\cdot \vec x}}\,\Psi (x) \
{}_{{\rm HQ}}\langle B_s, v|\overline{s}(x)\gamma^\mu\gamma_5 b_v(0)|0\rangle\
\ \ .}
Performing an operator product expansion
\eqn\oper{\bar s(x)\gamma^\mu\gamma_5 b_v(0) =
\bar s(0)\gamma^\mu\gamma_5 b_v(0) + x^k \partial_k \bar
s(0)\gamma^\mu\gamma_5
b_v(0) + ...\ \ ,}
and using
the definition
\eqn\fb{
\bra 0 \bar s \gamma^{\mu} \gamma_5 b \ket {B_s, v}_{{\rm HQ}}
\equiv i f_{B_s} m_{B_s} v^{\mu}/\sqrt{2 m_{B_s}},
}
of the $B_s$ meson decay constant gives
\eqn\form{\Omega_1 (\a \vec q\,) =
{1\over {\sqrt{2}}}f_{B_s}\sqrt{m_{B_s}}
\int d^3x\ e^{i{\vec q\cdot \vec x}}\,\Psi (x) \ \ \ ,}
where we have retained only the
first term in the operator product expansion, Eq.~\oper.
The above computation also applies to the
case where the final meson is $B_d$ instead of $B_s$, with the
replacement $f_{B_s}\sqrt{m_{B_s}}\rightarrow f_{B_d}\sqrt{m_{B_d}}$.  In
the $SU(3)_V$ limit,
$f_{B_d}\sqrt{m_{B_d}}=f_{B_s}\sqrt{m_{B_s}}$.
The leading $SU(3)_V$-violating correction to this
result can be found in ref.
\ref\gjmsw{B.~Grinstein, E.~Jenkins, A.V.~Manohar,~M.J. Savage
and M.B. Wise, UCSD/PTH 92-05 (1992).}.

In the $m_b\gg m_c\gg\lqcd$ limit, the wavefunction $\Psi(x)$ for the
$B_c$ is a Coulomb wavefunction, so that the form factor $\Omega_1$ can
be computed explicitly,
\eqn\omonecoul{
\Omega_1(\a \vec q) =
{1\over {\sqrt{2}}}f_{B_s}\sqrt{m_{B_s}} {8 \pi^{1/2} \a^{3/2}\over
(1 + \a^2\, {\vec q}^{\,2})^2}.
}
There will be corrections to the Coulomb form of the $B_c$ wavefunction,
because confinement effects are significant at the $c$
quark mass. Confinement effects in the meson wavefunction at the $c$
quark scale have been studied in detail in the $\psi$ system. A quark
model with a modified Coulomb potential provides a very good
description of the spectrum and radiative decays for the $\psi$. A
similar computation for the $B_c$ should provide a good description
of the $B_c$ wavefunction for use in Eq.~\form.

The $c$ and $b$ quark fields were treated as free fields in the
computation of Eq.~\form. Radiative gluon corrections
are in principle important. Gluon exchange between the $\bar b$ and $c$
quarks of the $B_c$ and between the $\bar b$ and $s$ quarks of the $B_s$
has already been included exactly in the definition of the states. The
only gluon contributions that are not included are radiative gluon
corrections where  gluons are exchanged between the $B_c$ and $B_s$.
This gluon exchange
leads to a violation of factorization in the computation of the
decay amplitude. The $B_c$ does not couple to gluons in the limit that
its radius becomes zero. For a finite radius $\a$,
the leading gluon coupling of the $B_c$ is to a two
gluon operator with coefficient proportional to $\a^3$. This produces a
small (and incalculable) correction to the decay form factors.

The decay $B_c\rightarrow D$ proceeds through the quark decay
$\bar b\rightarrow \bar u$. The $D$ meson is light compared
with the energy
released in the decay, and so can have a large recoil momentum. The
approximation methods used in this paper cannot be applied in this case.
However, there is a region of phase space near zero recoil where $\vec q
\ltap m_c$, where the heavy $c$ quark expansion is still valid. The
computation of the $\Sigma_1$ form factor in this region is almost
identical to the computation described above for $\Omega_1$, with the
result
\eqn\sigmaff{\Sigma_1 (\a \vec q\,) =
{1\over {\sqrt{2}}} f_{D}\sqrt{m_D}
\int d^3x\ e^{i{\vec q\cdot \vec x}}\,\Psi (x) \ \ \ .}
Eqs.~\form\ and \sigmaff\ both depend on the wavefunction of
the $B_c$ meson.  One can obtain a more reliable extraction
of KM mixing angles by considering the ratio $\Sigma_1(0)/
\Omega_1(0)$, which  should be  insensitive to
the detailed form of the $B_c$ wavefunction, and thus
provides a way of measuring $|V_{ub}|/|V_{cs}|$ and $|V_{ub}|/|V_{cd}|$.

The decay $B_c\rightarrow \eta_c$ proceeds through the quark decay
$\bar b\rightarrow \bar c$.
As for $B\rightarrow D$, the heavy quark expansion is
only valid in the region $\vec q \ltap m_c$ near the zero recoil point.
The form factor $\Delta$ is
calculable in terms of the wavefunctions of the
$\bar b c$ and $c \bar c$ bound states.
A straight-forward derivation yields
\eqn\del{
\Delta (\a \vec q\,) = 2 \int d^3 x  e^{-i{\vec q\cdot \vec x/2}}
\ \Psi_{\eta_c}^*(x) \Psi_{B_c} (x),
}
where the convolution of the two wavefunctions depends on the
radii
$\a$ and $a_\eta$ of the $B_c$ and $\eta_c$.
The details of this computation are nearly identical to
those found in ref.
\ref\white{M.J. White and M.J. Savage, \pl{271}{1991}{410}.}
for the semileptonic decay of baryons containing two heavy quarks.
In the limit that both states are described by a Coulomb wavefunction,
\eqn\deltacoul{
\Delta(\a\vec q\,)=16{\a^{3/2} a_\eta^{3/2}\over (\a+a_\eta)^3}
\left[1 + {{\vec q}^{\,2} \a^2 a_\eta^2\over 4
(\a+a_\eta)^2}\right]^{-2}.
}

\newsec{Corrections}

Corrections to the results of the previous sections can be divided
into two categories.  The first set of corrections are corrections
to heavy quark spin symmetry, and affect the relations
derived in Sect.~2.  The second set of corrections relate to the
validity of factorization and the
operator product expansion used in Sect.~4, as well as to the details of
the $B_c$ wavefunction.

Let us first consider corrections to heavy quark spin symmetry.
In this paper, we have worked in the limit $m_b\gg m_c
\gg \lqcd$.  There
are corrections to the heavy quark theory due to the finite
mass of the $b$ quark, which are of order $\lqcd/m_b$ and $m_c/m_b$. These
corrections are small and will not be discussed further.
In addition, there are
violations of the $c$ quark spin symmetry in the $B_c$. These arise
from interactions of the $c$ quark spin with the $b$ quark spin,
with the orbital
angular momentum of the $c$ quark, and with light degrees of freedom. The
$c$-$b$ spin-spin interaction is a $1/m_b$ effect, and is small. There is no
$c$ quark spin-orbit interaction for the $B_c$ because the $c$ quark is in an
$s$-wave. At lowest order, the $B_c$ is made up of a $\bar b$
quark and $c$ quark in
a bound state. There are, however, corrections to this form in which the
$B_c$ wavefunction
also
contains additional light degrees of freedom. In a bag model, this
would correspond to exciting gluonic excitations in the bag. The spin coupling
of the $c$ quark to these light degrees of freedom violates the $c$ quark spin
symmetry. The interaction energy is of order $\lqcd^2/m_c$. Since the energy
cost of exciting a light degree of freedom is of order $\lqcd$, the net
spin symmetry violation in the matrix element is of order $\lqcd/m_c$.

The results of Sect.~4 depend not only on taking $m_c\gg \lqcd$, but
also on the size of the $B_c$ being much smaller than
$\lqcd^{-1}$.
The higher derivative terms in the operator product expansion, Eq.~\oper\
produce corrections of order $\a \lqcd$, because each factor of
$\partial$ on the light quark operator produces a factor of $\lqcd$ in the
matrix element, and each factor of $x$ produces a factor of the size of the
bound state $\a$. There are also violations of factorization in the
operator matrix element Eq.~\mint. As discussed in the previous section,
gluon interactions with the $B_c$ are of order $\a^3$.
There is no suppression factor
for the interaction of gluons with the $B_s$, since the $B_s$ has a size of
order $\lqcd$. Thus the gluon interactions produce corrections of order $(\lqcd
\a)^3$. For the $B_c\rightarrow \eta_c, \psi$ decay, there is an additional
suppression factor of $a_\eta^3$ for the interaction of gluons with the
$\eta_c, \psi$, so that  the net correction is of order $\lqcd^6\a^3a_\eta^3$.
There are non-perturbative corrections to the Coulomb wavefunction of
the $B_c$, which are of order $\lqcd/m_c$. As discussed in the previous
section, most of these effects can be included by modeling the $B_c$ by
a realistic potential which is adjusted to correctly reproduce the $B_c$
excitation spectrum.
There are also radiative corrections which produce corrections of order
$\alpha_s(m_b)$ and $\alpha_s(m_c)$.
Finally, there are $1/m_c$ recoil corrections for $B_c\rightarrow D$ and
$B_c\rightarrow \eta_c, \psi$ which are of order $\vec q/m_c$.

\newsec{Conclusions}

The semileptonic decay $B_c\rightarrow D\ell\nu$ provides
a way of extracting the weak
mixing angle $\abs{V_{ub}}$. Theoretical uncertainties can be minimized by
extracting the ratio $\abs{V_{ub}/V_{cs}}$ using the ratio of the
$B_c\rightarrow D$ and $B_c\rightarrow B_s$ form factors near zero
recoil. The number of $B_c$'s produced in hadron collisions is much
smaller than the number of $B$'s.
Nevertheless, the $B_c$ meson still
provides an alternative measurement of $\abs{V_{ub}}$ to the
value which will be obtained
by comparing  semileptonic
$B\rightarrow \rho\ell\nu$ and $D\rightarrow \rho\ell\nu$ decays. Both
of these extractions will have corrections due to the finite mass of
the $c$ quark whose numerical importance will have to be determined
experimentally.

\bigskip
\centerline{\bf Acknowledgements}
A.M. would like to thank E.~Eichten and C.~Quigg for an interesting
discussion on the $B_c$ which started this investigation. Eichten and
Quigg have computed the spectrum and radiative decays in the $\bar b c$
sector using a potential model \ref\eichten{E.~Eichten and C.~Quigg,
private
communication.}. Some aspects of $B_c$ decays have also been
computed previously by M.~Adler \ref\adler{M. Adler, Caltech Ph.D. Thesis 1989,
(unpublished).}.
This work was supported in part by DOE grant \#DE-FG03-90ER40546,
and by a NSF
Presidential Young Investigator award \pyiam.
\bigskip

\listrefs
\end